\documentclass[aps,prb,superscriptaddress,preprint,showpacs]{revtex4-1}

	\usepackage{amsmath}
	\usepackage{amssymb} 
	\usepackage{color}
	\usepackage{makeidx}
	\usepackage{amsfonts}
	\usepackage[ansinew]{inputenc}
	\usepackage[usenames,dvipsnames]{pstricks}
	\usepackage{subfigure}
	\usepackage{epsfig}
	\usepackage{pst-grad} 
	\usepackage{pst-plot} 
	\usepackage[colorlinks,hyperindex]{hyperref}

	\hypersetup
	{
		colorlinks,%
		citecolor=black,%
		linkcolor=black,%
		urlcolor=black,%
	}

\makeindex

\begin{document}

\title{The thermal conductivity of silicon nitride membranes is not sensitive to stress}

\author{Hossein Ftouni}
\affiliation{Institut N\'EEL, CNRS, 25 avenue des Martyrs, F-38042 Grenoble, France}
\affiliation{Univ. Grenoble Alpes, Inst NEEL, F-38042 Grenoble, France}

\author{Christophe Blanc}
\affiliation{Institut N\'EEL, CNRS, 25 avenue des Martyrs, F-38042 Grenoble, France}
\affiliation{Univ. Grenoble Alpes, Inst NEEL, F-38042 Grenoble, France}

\author{Dimitri Tainoff}
\affiliation{Institut N\'EEL, CNRS, 25 avenue des Martyrs, F-38042 Grenoble, France}
\affiliation{Univ. Grenoble Alpes, Inst NEEL, F-38042 Grenoble, France}

\author{Andrew D. Fefferman}
\affiliation{Institut N\'EEL, CNRS, 25 avenue des Martyrs, F-38042 Grenoble, France}
\affiliation{Univ. Grenoble Alpes, Inst NEEL, F-38042 Grenoble, France}

\author{Martial Defoort}
\affiliation{Institut N\'EEL, CNRS, 25 avenue des Martyrs, F-38042 Grenoble, France}
\affiliation{Univ. Grenoble Alpes, Inst NEEL, F-38042 Grenoble, France}

\author{Kunal J. Lulla}
\affiliation{Institut N\'EEL, CNRS, 25 avenue des Martyrs, F-38042 Grenoble, France}
\affiliation{Univ. Grenoble Alpes, Inst NEEL, F-38042 Grenoble, France}

\author{Jacques Richard}
\affiliation{Institut N\'EEL, CNRS, 25 avenue des Martyrs, F-38042 Grenoble, France}
\affiliation{Univ. Grenoble Alpes, Inst NEEL, F-38042 Grenoble, France}

\author{Eddy Collin}
\affiliation{Institut N\'EEL, CNRS, 25 avenue des Martyrs, F-38042 Grenoble, France}
\affiliation{Univ. Grenoble Alpes, Inst NEEL, F-38042 Grenoble, France}

\author{Olivier Bourgeois}
\affiliation{Institut N\'EEL, CNRS, 25 avenue des Martyrs, F-38042 Grenoble, France}
\affiliation{Univ. Grenoble Alpes, Inst NEEL, F-38042 Grenoble, France}
\email{olivier.bourgeois@neel.cnrs.fr}

\pacs{65.60.+a,68.65.-k,62.40.+i,63.50.Lm}

\begin{abstract}

We have measured the thermal properties of suspended membranes from 10~K to 300~K for two amplitudes of internal stress (about 0.1~GPa and 1~GPa) and for two different thicknesses (50~nm and 100~nm). The use of the original 3$\omega$-Volklein method has allowed the extraction of both the specific heat and the thermal conductivity of each SiN membrane over a wide temperature range. 
The mechanical properties of the same substrates have been measured at helium temperatures using nanomechanical techniques.
Our measurements show that the thermal transport in freestanding SiN membranes is not affected by the presence of internal stress. Consistently, mechanical dissipation is also unaffected even though $Q$s increase with increasing tensile stress.
We thus demonstrate that the theory developed by Wu and Yu [Phys. Rev. B, \textbf{84}, 174109 (2011)] does not apply to this amorphous material in this stress range. On the other hand, our results can be viewed as a natural consequence of the "dissipation dilution" argument [Y. L. Huang and P. R. Saulson, Rev. Sci. Instrum. \textbf{69}, 544 (1998)] which has been introduced in the context of mechanical damping.

\end{abstract}

\maketitle

\section{Introduction}

Silicon nitride (SiN) thin films are widely used to thermally isolate sensitive thermal detectors, for etch masking as well as layers for  micro-electromechanical systems \cite{Garimella2006}. Indeed, outstanding mechanical properties including very high quality factors $Q$\cite{Southworth2009,Defoort2013a} can be reached in optimized SiN material. Depending on deposition parameters, SiN films can experience very large residual (biaxial) stress during deposition. It is thus of prime importance to understand the role of the internal stress not only on the mechanical properties but also on the other physical characteristics of SiN films including optical, thermal and electrical properties. Silicon nitride has a specific place due to its amorphous nature and the study of stress in that compound is also an issue for the fundamental understanding of its role in the physics of glasses \cite{Southworth2009}.

Using the stress to tune the thermal properties of nanomaterials is one of the possible ways to design future thermal components (thermal rectifier, thermal diode, thermal switch \cite{Li2012} ...). This has been proposed for monocrystalline silicon \cite{Li2010,Paul2011}, as strain in silicon is currently used to enhance electron mobility in transistors \cite{Mii1991}. Since the debate on the origin of mechanical dissipation in strained glasses like SiN \cite{Southworth2009,NanoLett2007} the question of the effect of stress, whatever its origin (internal or external) on the thermal properties has been raised and theoretically addressed for the case of silicon nitride \cite{Wu2011}. Indeed it is well known that stoichiometric silicon nitride (Si$_3$N$_4$) prepared by low pressure chemical vapor deposition (LPCVD) contains a significant internal tensile stress (up to about 1 GPa) as compared to regular non-stoichiometric SiN that has a very low internal stress (below 0.2 GPa).
In an attempt to explain the very high mechanical $Q$s, Wu and Yu \cite{Wu2011} proposed a model where the internal losses in the material are sensitive to the stress state. Their calculations based on this hypothesis predict that the thermal conductivity of SiN may be strongly enhanced by the presence of stress. 
On the other hand, systematic mechanical measurements on high stress SiN substrates explain the $Q$s through the so called "dissipation dilution" model \cite{Huang1998}: mechanical energy is stored through the tensioning of the substrate while the dissipation is unaffected \cite{Unter2010,Yu2012,Wu2011}. However, no experiments to date compared directly similar devices made of different SiN materials.
Furthermore, thermal properties of low stress SiN have been widely measured over a broad temperature range \cite{Leivo1998,Holmes1998,Zink2004,Queen2009,Sultan2013} for different kinds of thin films and nanomaterials, but very few experimental studies deal with the influence of stress on the thermal transport at the nanoscale \cite{Alam2012}.

In order to study the potential effect of internal stress on the thermal properties of silicon nitride, the thermal conductivity and the specific heat have been measured as a function of temperature for high stress (HS) and low stress (LS)  SiN membranes having a thickness of 50~nm and 100~nm. These measurements are performed using the 3$\omega$-V$\ddot{\rm{o}}$lklein method \cite{Sikora2012,Sikora2013,Ftouni2013} as described in previous papers. This technique allows the measurement of both thermal conductivity and specific heat of a given membrane within the same experiment over a broad temperature range.
The mechanical dissipation and stress amplitude of both HS and LS substrates are also measured at cryogenic temperatures by means of nanomechanical resonators \cite{Defoort2013b}. We show experimentally that thermal conduction is essentially independent of the stress stored in this material. This is inconsistent with the hypothesis underlying the model of Ref.~\cite{Wu2011} for SiN, and corroborates the "dissipation dilution" explanation for high mechanical $Q$s.

\section{Samples and experimental methods}

\begin{figure}[ht]
	\begin{center}
		\includegraphics[width=7cm]{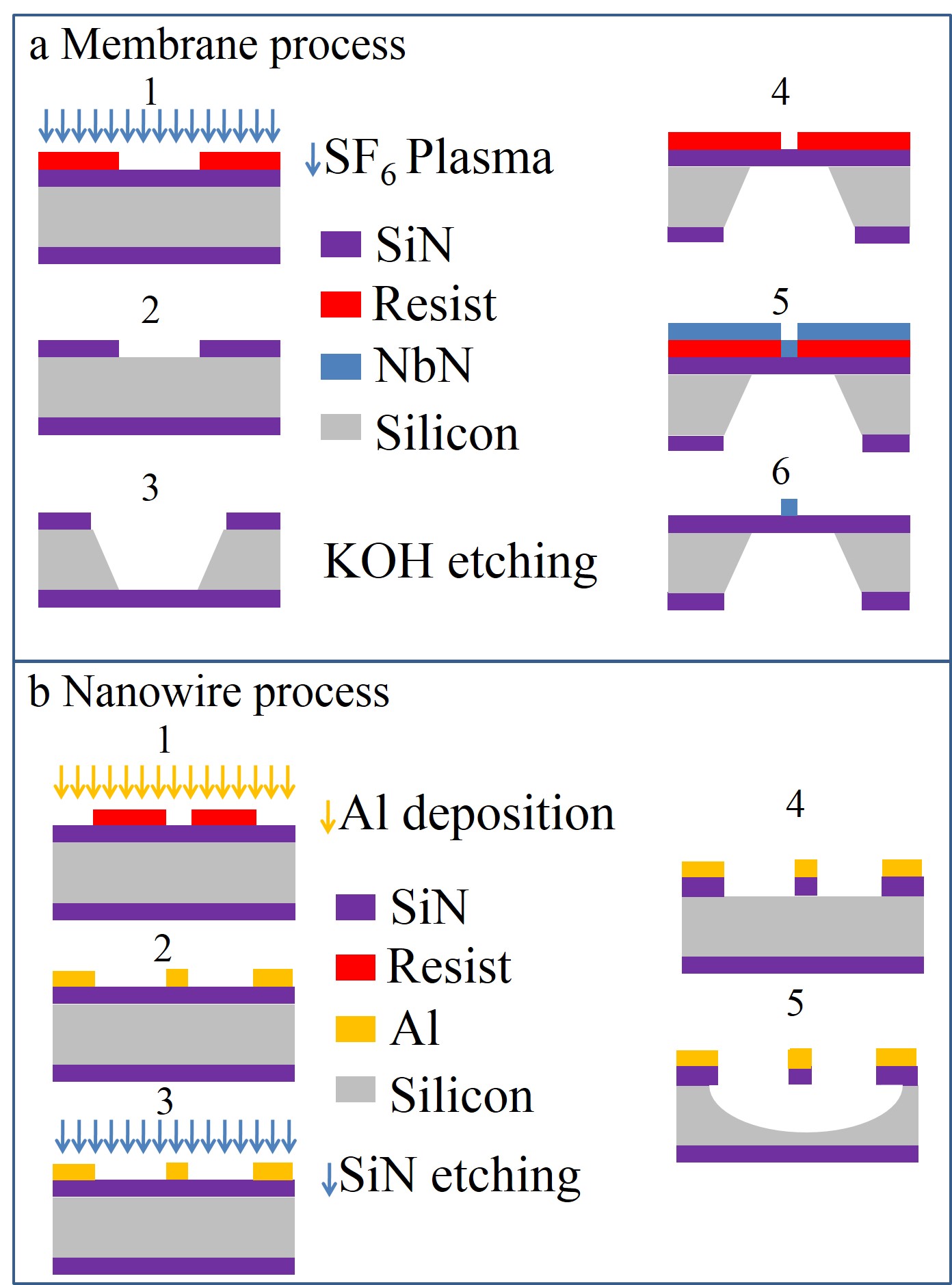} 
		\caption{(colour online) Micro and nanofabrication processes of both suspended structures studied in this work. a) fabrication process for the membrane 1), 2) The patterns of the membranes are created by photolithography. The non protected SiN is removed by SF$_{6}$ RIE etching. 3) The silicon is anisotropically etched in a KOH solution. 4) The thermometers are obtained by a lift-off process; the area is patterned by photolithography. 5) NbN (70~nm) is deposited by reactive sputtering. 6) The resist and NbN layer is removed using a wet procedure. b) fabrication process for the nanowires 1) The patterns of the nanowires are created by e-beam lithography. 2) evaporation of Al layer (30~nm) 3), 4) The non-protected SiN is removed by SF$_{6}$ RIE. 5) The silicon is isotropically etched by gazeous XeF$_2$ ecthing.}
		\label{fab}
	\end{center}
\end{figure}

The thermal properties of two types of SiN membranes have been measured: high stress stoichiometric Si$_3$N$_4$ and low stress SiN deposited by LPCVD. The amorphous stoichiometric high stress (HS) Si$_3$N$_4$ as well as low stress (LS) SiN were grown on both sides of a silicon substrate. The membranes were then patterned on the rear side by laser photolithography. After removing the silicon nitride by SF$_6$ Reactive Ion Etching, the silicon substrate on the rear side was etched in KOH, as described in Fig.~\ref{fab}. The final result is a rectangular SiN membrane obtained on the front side.

\begin{figure*}[ht]
	\begin{center}
		\includegraphics[width=16cm]{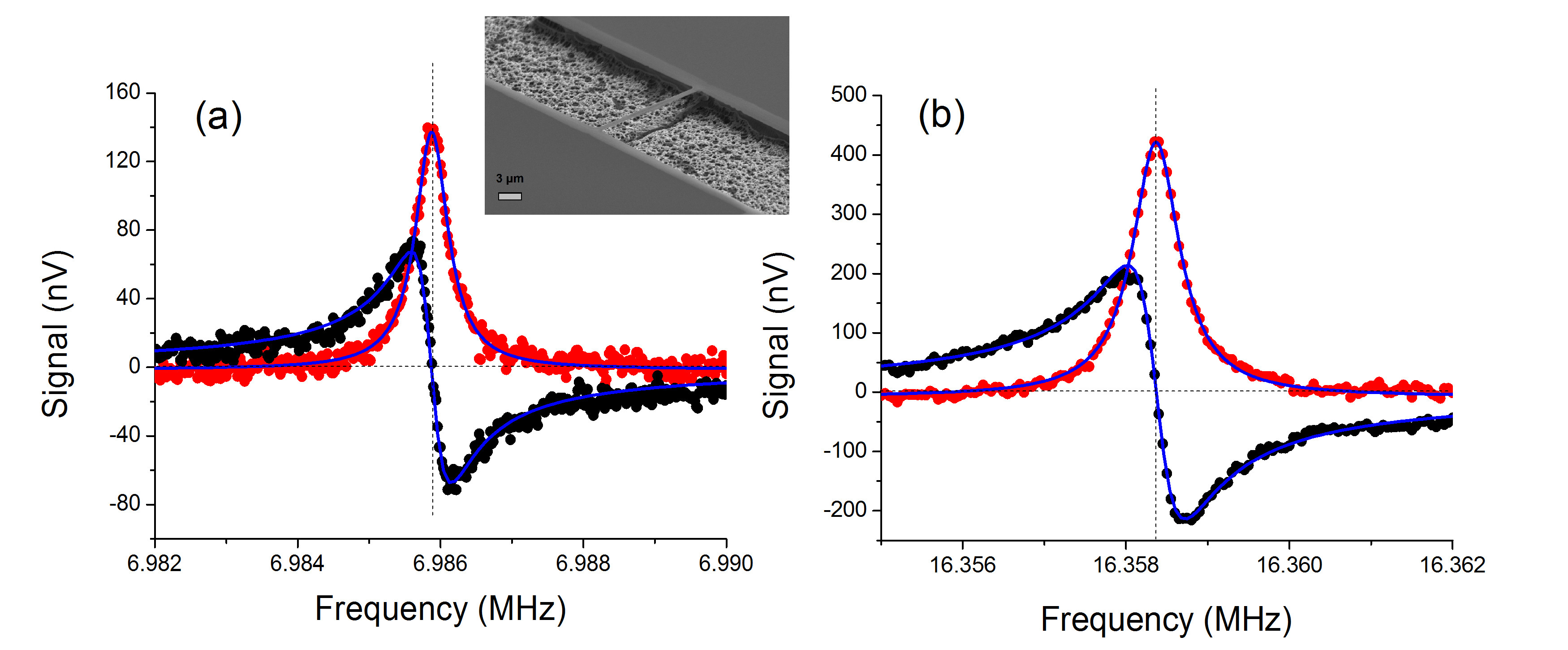} 
		\caption{(colour online) Measurement of the first flexure resonance of two suspended SiN beams with different in-built stresses, the red points are the in-phase signal and the black points the out-of-phase signal (both 15~$\mu$m long, 250~nm wide, 100~nm thick, see SEM picture in inset). (a) low-stress and (b) high-stress resonance lines obtained in the linear regime. The lines are Lorentzian fits, with full width at half height of $650 \pm 50~$Hz (HS, $Q\approx 25\,000$) and $500 \pm 50~$Hz (LS, $Q\approx 14\,000$). We extract from the resonance frequencies the stress values of $0.85 \pm 0.08$ GPa (HS) and $0.12 \pm 0.05$ GPa (LS). Data taken at 4.2$~$K in vacuum.}
		\label{stress_meas}
	\end{center}
\end{figure*}

Before the thermal study, a mechanical measurement was performed to quantify the stress present after releasing the membranes. A suspended silicon nitride beam with 100~nm thickness, 250~nm width and 15~$\mu$m length fabricated using e-beam lithography from the same substrate was placed in a magnetic field (see Fig.~\ref{fab} for fabrication details). A sinusoidal driving current within a 30~nm thin deposited Al layer is used to generate the Lorentz force causing the beam's out-of-plane oscillation \cite{Cleland1999}. This measurement is performed in a vacuum of about 10$^{-6}$~mbar at helium temperatures. The magnetic flux cut by the beam oscillation generates a voltage which is measured using a lock-in amplifier \cite{Collin2012,Defoort2013a}. Typical resonance curves for the first flexure and their respective fits are shown in Fig.~\ref{stress_meas}. 

The expression for the $n^{th}$ mode resonance frequency of a stressed doubly-clamped beam is given by \cite{Parpia2007}:
\begin{equation}
f_n=\frac{n}{2} \sqrt{\frac{\sigma}{\rho h^2}}
\label{stress_eq}
\end{equation}

Eq. \ref{stress_eq} is used to calculate the stress $\sigma$ within the beam, with $n$ the mode number, $\rho$ the silicon nitride density (3$~$g/cm$^3$) and $h$ the beam length \cite{Defoort2013b}. We find a stress value of about 0.85~GPa for the HS silicon nitride, which confirms that the membrane is still stressed after releasing and 0.12~GPa for the LS silicon nitride. 
Note that both values fall in the high-stress limit of beam theory, validating the use of Eq.~(\ref{stress_eq}). At the same time, the linewidth of the resonances in Fig.~\ref{stress_meas} (measuring mechanical dissipation) are almost equal for HS and LS beams. Even though the stress is modified by almost an order of magnitude, mechanical friction around 4 Kelvin seems insensitive. 
Many devices varying shapes and stress have been measured, studying flexural modes up to $n = 9$ and confirming these findings \cite{MartialPhD}.
The temperature dependence of mechanical properties shall be discussed elsewhere \cite{ToCome}.

These mechanical measurements seem to be in contradiction with the claims of Ref.~\cite{Southworth2009} (in which very different types of devices were compared),  
but the raw data do agree: we find indeed a higher $Q$ for a higher stress \cite{Southworth2009,NanoLett2007}. On the other hand, our results are a natural consequence of the "dissipation dilution" model \cite{Huang1998,Unter2010,Yu2012} in which high $Q$s are reached because of the flexural energy stored in the tensioning of the structures: while in Refs.~\cite{Unter2010,Yu2012} only HS devices are measured, we show here with a comparison between geometrically identical HS and LS structures that there is no internal effect of stress on mechanical dissipation.

Thermal experiments are conducted on the very same materials. As mentioned above, we have chosen in this study the most appropriate method to measure the thermal conductivity of large aspect ratio suspended membranes: the 3$\omega$-V$\ddot{\rm{o}}$lklein method \cite{Sikora2012,Sikora2013,Ftouni2013}.
The principle of the method consists in creating a sinusoidal Joule heating generated by an AC electrical current at frequency $\omega$ across a transducer centered along the long axis of a rectangular membrane. The center of the membrane is thermally isolated from the frame and hence its temperature is free to increase. 

The temperature oscillation ($\approx$100~mK) of the membrane is at 2$\omega$ and is directly related to its thermal properties by the amplitude and the frequency dependence of the aforementioned temperature oscillation. Since the resistance of the thermometer can be considered as linearly dependent on temperature over that small temperature oscillation, the voltage $V=R[T(2\omega)]\times I(\omega)$ will have an ohmic component at $\omega$ and a thermal component at 3$\omega$. By measuring the V$_{3\omega}$ voltage appearing across the transducer as function of the frequency, it is possible to deduce the thermal conductivity and the specific heat of the membrane \cite{Ftouni2013}. The membranes measured in this study are 300~$\mu$m wide and 1.5~mm long.
 
\begin{figure}[ht]
	\begin{center}
		\includegraphics[width=7cm]{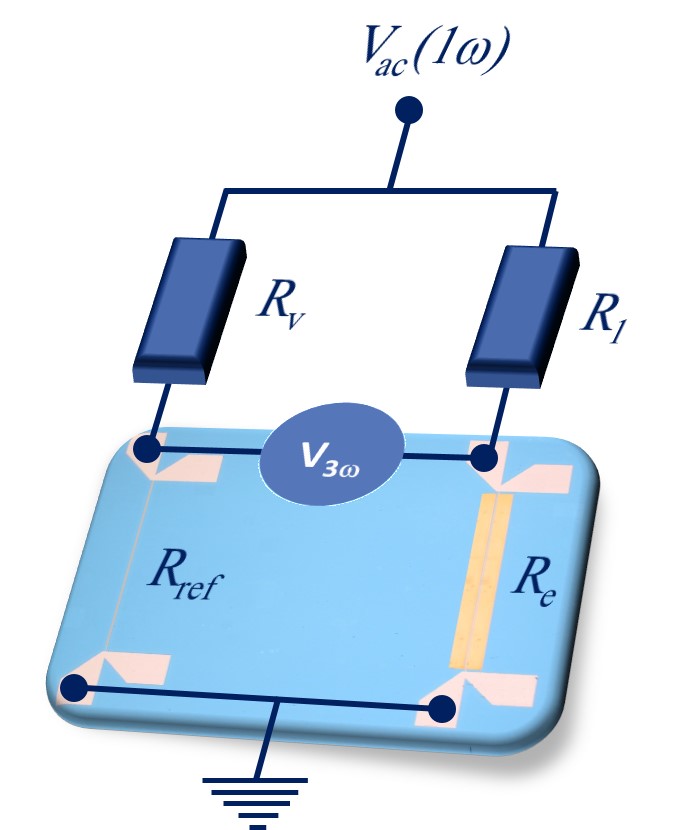} 
		\caption{(colour online) Experimental set-up based on the Wheatstone bridge configuration, the yellow membrane sample is on the bottom right and the reference thermometer is on the left; the blue area of the sketch corresponds to the temperature regulated part of the Wheatstone bridge. }
		\label{whb}
	\end{center}
\end{figure}

The transducer of 5~$\mu$m width and 1.5~mm length is made out of NbN whose resistance is strongly temperature dependent. It serves as a thermometer and a heater at the same time \cite{Bourgeois2006,Heron2009a,Lopeandia2010}. For the present work, the thermometer has been designed for the 10~K to 320~K temperature range. Typically, the resistance of the thermometer is about 100~kOhm at room temperature with a temperature coefficient of resistance (TCR) $\alpha=\frac{dR}{RdT}$ of 10$^{-2}$ K$^{-1}$ at 300~K and of 0.1~K$^{-1}$ at 4~K.

Since the 1$\omega$ voltage is 3 to 4 orders of magnitude higher than the 3$\omega$ voltage, a specific Wheatstone bridge is used to reduce the 1$\omega$ component and perform thermal measurements (see Fig.~\ref{whb}). The bridge consists of the measured sample with a resistance R$_e$, which is the NbN thermometer on the SiN membrane, the reference thermometer R$_{\rm{ref}}$ deposited on the bulk region of the chip which has the same geometry and deposited in the same run as the transducer on the membrane, an adjustable resistor R$_v$, and an equivalent nonadjustable resistance R$_1$ =50~kOhm.

The general expression of the measured 3$\omega$ output Wheatstone bridge voltage can be given by \cite{Sikora2012,Sikora2013}:

\begin{equation}
\left|V_{3\omega}^{rms}(\omega)\right|= \frac{V^{rms}_{ac}\alpha R{e}R{_1}\left|\Delta T_{2\omega} \right|}{2(R_e + R_1){^2}  }
\label{V3w_dt}
\end{equation}

with $\alpha$ the TCR, $V_{ac}$ the 1$\omega$ input Wheatstone bridge voltage and $\left|\Delta T_{2\omega} \right|$ the amplitude of the temperature oscillation at 2$\omega$ of the membrane due to the sinusoidal nature of heating.

By solving the partial differential equation of the heat flux across the membrane, eq.~\ref{V3w} gives the relation between the thermal properties, the dimensions of the membrane and $V_{3\omega}$ \cite{Sikora2012,Sikora2013}:

\begin{equation}
\left|V_{3\omega}^{rms}(\omega)\right|=\frac{\alpha  (V_{ac}^{rms})^{3}R_{1}R_{e}^{2}}{4K_{p}\left(R_{e}+R_{1}\right)^{4}\left[1+\omega^{2}\left(4\tau^{2}+\frac{2{\ell}^{4}}{3D^{2}}+\frac{4\tau {\ell}^{2}}{3D}\right)\right]^{1/2}}
\label{V3w} 
\end{equation}

with $K_{p}=\frac{kS}{\ell}$ the thermal conductance and $C=cS\ell$ the heat capacity of the measured membrane, $\tau=\frac{C}{K_{p}}$ the thermalization time of the membrane to the heat bath, $D=\frac{k}{\rho c}$ the thermal diffusivity, ${\ell}$ half the width of the membrane and $S$ the section of the membrane (perpendicular to the heat flow). By measuring the $V_{3\omega}$ voltage as a function of frequency both $k$ (in-plane thermal conductivity) and $c$ (specific heat) of the membrane can be extracted \cite{Ftouni2013}.

\begin{figure}[htbp]
	\begin{center}
		\includegraphics[width=8.5cm]{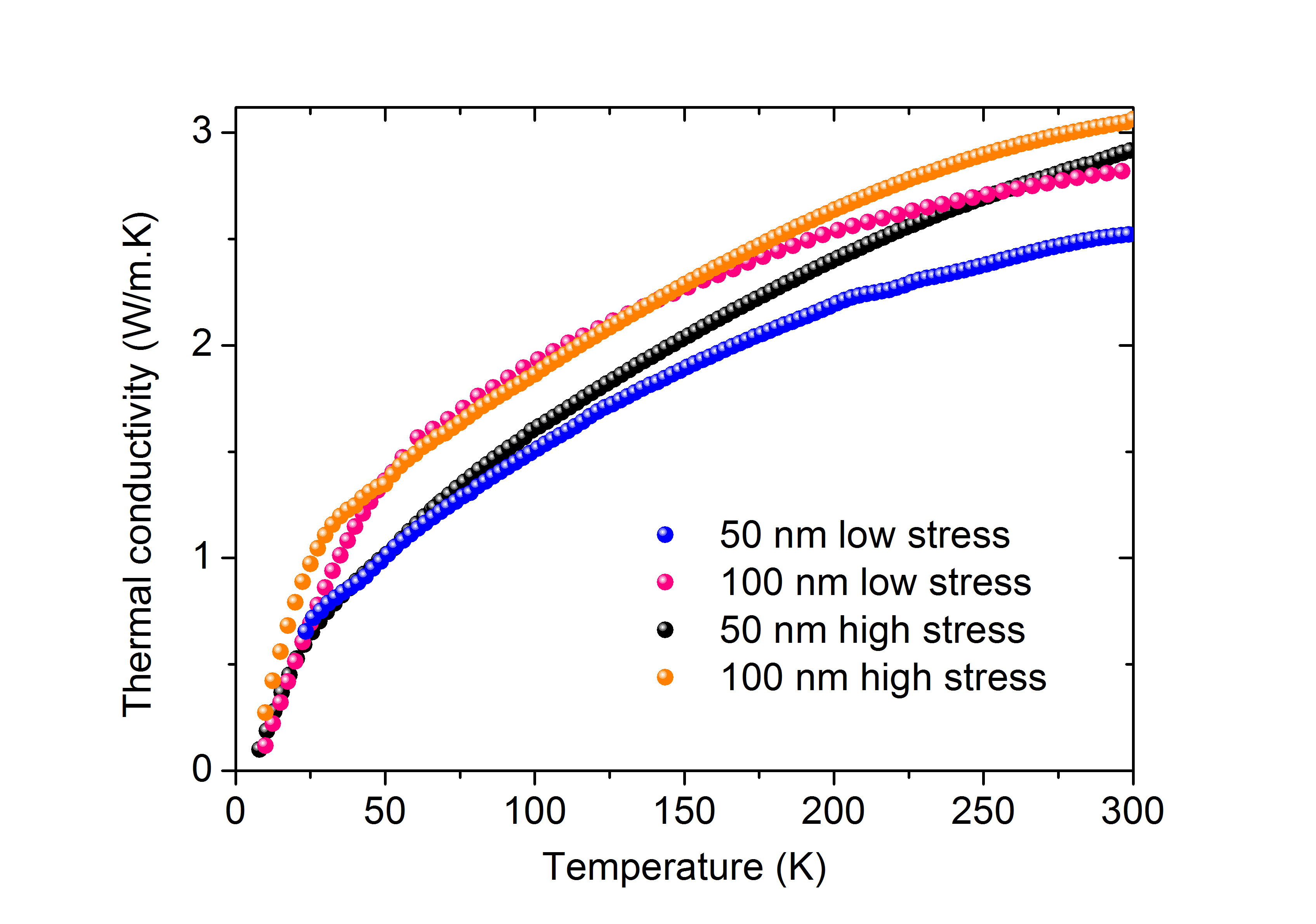} 
		\caption{(colour online) Thermal conductivity measurement of 50~nm and 100~nm thick membranes for both SiN low stress and high stress. The 100~nm curves of low stress and high stress show nearly no difference.}
		\label{k}
	\end{center}
\end{figure}

\section{Experimental results}

The thermal conductivity of the four different membranes (50 and 100~nm, low stress and high stress) has been measured versus temperature from 10~K to 300~K. The experimental data of thermal conductivity are presented in the Fig. \ref{k}. As expected for amorphous materials the thermal conductivity of all membranes is continuously increasing with temperature as observed by Queen and Hellman \cite{Queen2009}. The general trend of the temperature variation of thermal conductivity of all different SiN membranes (LS and HS) are very similar. Only the 50~nm LS membrane seems to have a slightly lower thermal conductivity at room temperature with a value approaching 2.5~W.m$^{-1}$.K$^{-1}$, instead of 3~W.m$^{-1}$.K$^{-1}$ for the others. 

In all cases, values of the thermal conductivity at room temperature are approximately 3~W.m$^{-1}$.K$^{-1}$. These values are in accordance with most of the in plane values of thermal conductivity measured which are displayed in table~\ref{tab:SiN}. Indeed Jain and Goodson \cite{Jain2008} have measured the in-plane thermal conductivity of 1.5~$\mu$m thick SiN specimens to be about 5~W.m$^{-1}$.K$^{-1}$. At the nanoscale, Sultan \textit{et al.} \cite{Sultan2009} reported thermal conductivity of 500~nm thin films as 3-4~W.m$^{-1}$.K$^{-1}$ for a temperature range of 77-325~K. For 180-220 nm thick LS nitride, Zink and Hellman \cite{Zink2004} also observed temperature variation of thermal conductivity ranging from 0.07 to 4~W.m$^{-1}$.K$^{-1}$ from 3 to 300~K. The cross-plane thermal conductivity measured by Lee and Cahill \cite{Lee1997} for less than 100~nm thickness was in the range of 0.4-0.7~W.m$^{-1}$.K$^{-1}$ showing severely reduced thermal conductivity, which was ascribed to the interfacial thermal resistance. Zhang and Grigoropuolos \cite{Zhang1995} also observed anomalous thickness dependence and suggested that micro structural defects may strongly influence thermal conductivity. It is important to note that none of the above studies measure thermal conductivity as a function of the internal stress. 

\begin{table}
\begin{ruledtabular}
\begin{center}
\scriptsize
\begin{tabular}{|c|c|c|c|c|c|c|}
  \hline
  Reference 	& Deposition  	& Stoichiometry 		& Stress 			& k at 300 K 		& c at 300 K 			& Sample \\
  \hline
  \cite{Mastrangelo1990} 	& LPCVD 		& Si$_{0.66}$N$_{0.34}$	 & not measured 				& 3.2 						& 0.7 						& free st. in- plane  \\
  \cite{Zhang1995}		 & LPCVD		 & Si$_1$N$_{1.1}$ 		& not measured 		 & 8-10 					& not measured				& free st. out of plane \\
  \cite{Jain2008} 		& LPCVD 		& Si rich 			& low stress			 & 4.5 					& 0.5 						&  free st. in-plane \\
  \cite{Sultan2009}		 & LPCVD 		& Si rich 			& low stress 			& 3.5						 &not measured  				& free st. in-plane \\
  \cite{Lee1997}		 & PECVD-APCVD 	& Si$_1$N$_{1.1}$		 & not measured		& 0.3 						& not measured				& out of plane \\
  \cite{Verbridge2006} 	& LPCVD 		&  not measured 		& high stress			 & 3.2 					& not measured 				& free st. in-plane \\
  \cite{Alam2012} 		& LPCVD 		&   Si$_1$N$_{1.1}$		 & from 0 to 2.4\% 		&  2.7 (LS) to 0.4 (HS) 	& not measured				 & free st. in-plane \\
  this work LS 			& LPCVD		& Si$_1$N$_{1.1}$		 &  0.2  GPa	&   2.5 				&  0.8						 & free st. in-plane \\
	  this work  HS			& LPCVD		& Si$_3$N$_4$			 & 0.85 GPa 	&   3 					&  0.8						 & free st. in-plane \\
  \hline
\end{tabular}
\caption{\label{tab:SiN} \small{Measured values of thermal conductivity ($k$ in W.m$^{-1}$.K$^{-1}$) and specific heat ($c$ in J/g.K) of silicon nitride having different stoichiometry and/or different stress. Our results are in accordance with most of the studies. LS is for Low Stress, HS for High Stress and free st. for free standing membranes.}}
\end{center}
\end{ruledtabular}
\end{table}

\begin{figure}[htb]
	\begin{center}
		\includegraphics[width=8.5cm]{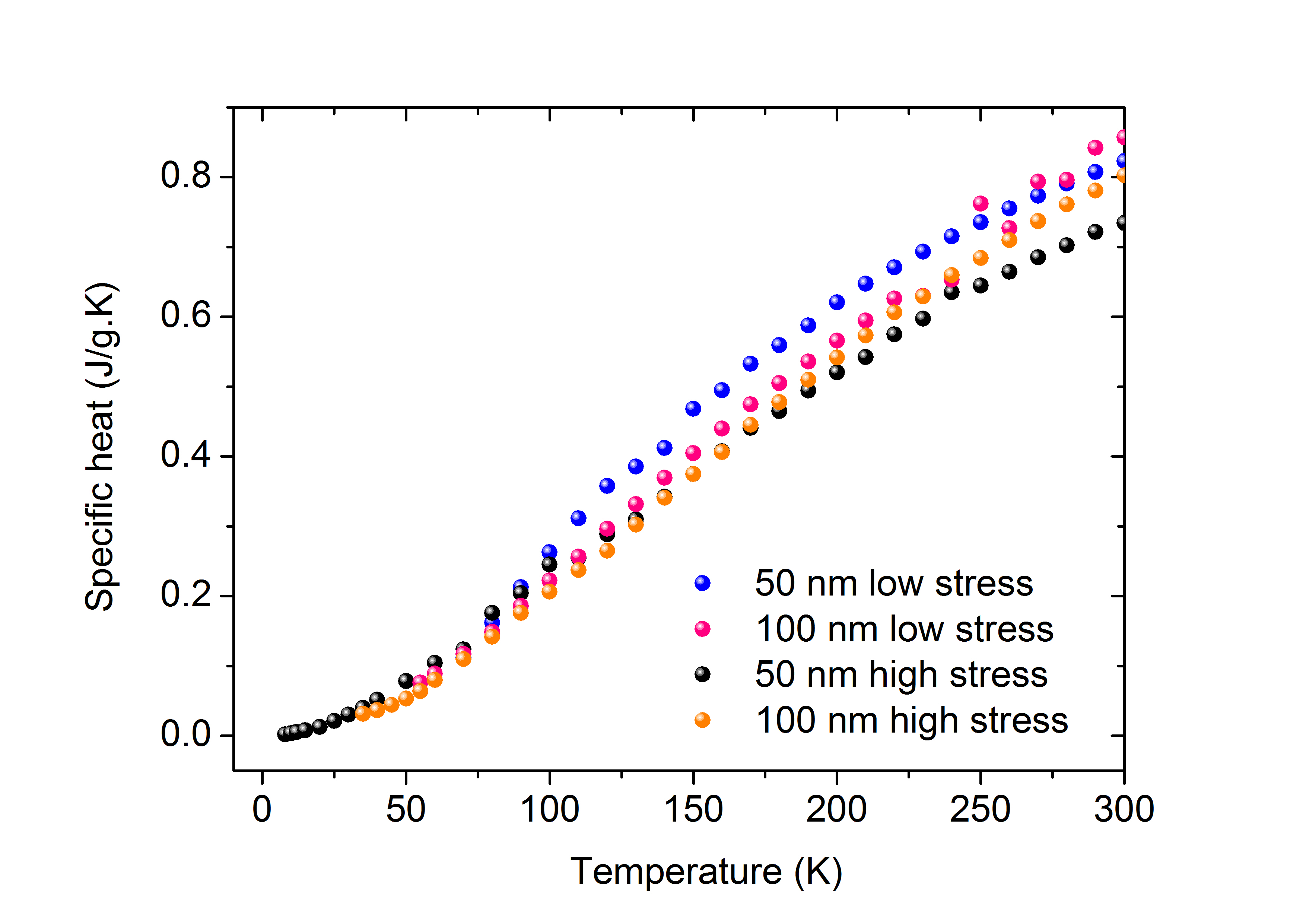} 
		\caption{(colour online) Specific heat measurement of 50~nm and 100~nm thick of both LS and HS silicon nitride from low temperature (10~K) to room temperature. }
		\label{c}
	\end{center}
\end{figure}

In order to verify the coherence of our experimental results, we have extracted the specific heat from the variation of the 3$\omega$ signal versus the frequency. Generally the specific heat is not expected to vary strongly as a function of stress at room temperature \cite{Wu2011}, and consequently it is a good test for the experiment. The results for the four different membranes are shown in Fig.~\ref{c}. The temperature variation of the specific heat is very similar for the four samples. For both 50 and 100~nm thick membranes we observe that the specific heat tends to be slightly higher for the case of low stress sample. But here again, the differences are insignificant and the specific heat is very similar for all the thicknesses and stress (low and high). The Debye temperatures deduced from the heat capacity measurements vary from 620 to 650~K depending on the sample which is a little lower than the commonly accepted value \cite{Zink2004}. Our measurements of thermal conductivity and specific heat demonstrate that no significant differences occur for the thermal transport in high and low stress SiN material because even with a stress close to 1~GPa, no modification of the phonon thermal conductivity can be observed.

\section{Discussion}

High stress silicon nitride mechanical devices exhibit remarkable $Q$ factors: inverse quality factors $Q^{-1}$  are two to three orders of magnitude lower than those of amorphous SiO$_2$ from 4~K up to room temperature \cite{Southworth2009}. The true origin of mechanical dissipation in stressed SiN is still unknown but could have connections with the thermal properties \cite{Southworth2009,Unter2010,Yu2012}. 
Even though amorphous solids are by nature diverse in composition, 
these materials are characterized by a universal behaviour of the thermal conductivity and mechanical dissipation at low temperature (between 0.1 and 10~K) \cite{Pohl2002,Zeller1971}. This universal behavior was initially reported by Zeller and Pohl \cite{Zeller1971} and  described in terms of a phenomenological model which takes into account the contribution from defects referred to as two-level systems (TLS) \cite{Phillips1972,Anderson1972}. 
The model does reproduce the data, but the universality appears as a surprising coincidence which continues to puzzle physicists \cite{Southworth2009,vural2011}.

\begin{figure}[htbp]
	\begin{center}
		\includegraphics[width=10cm]{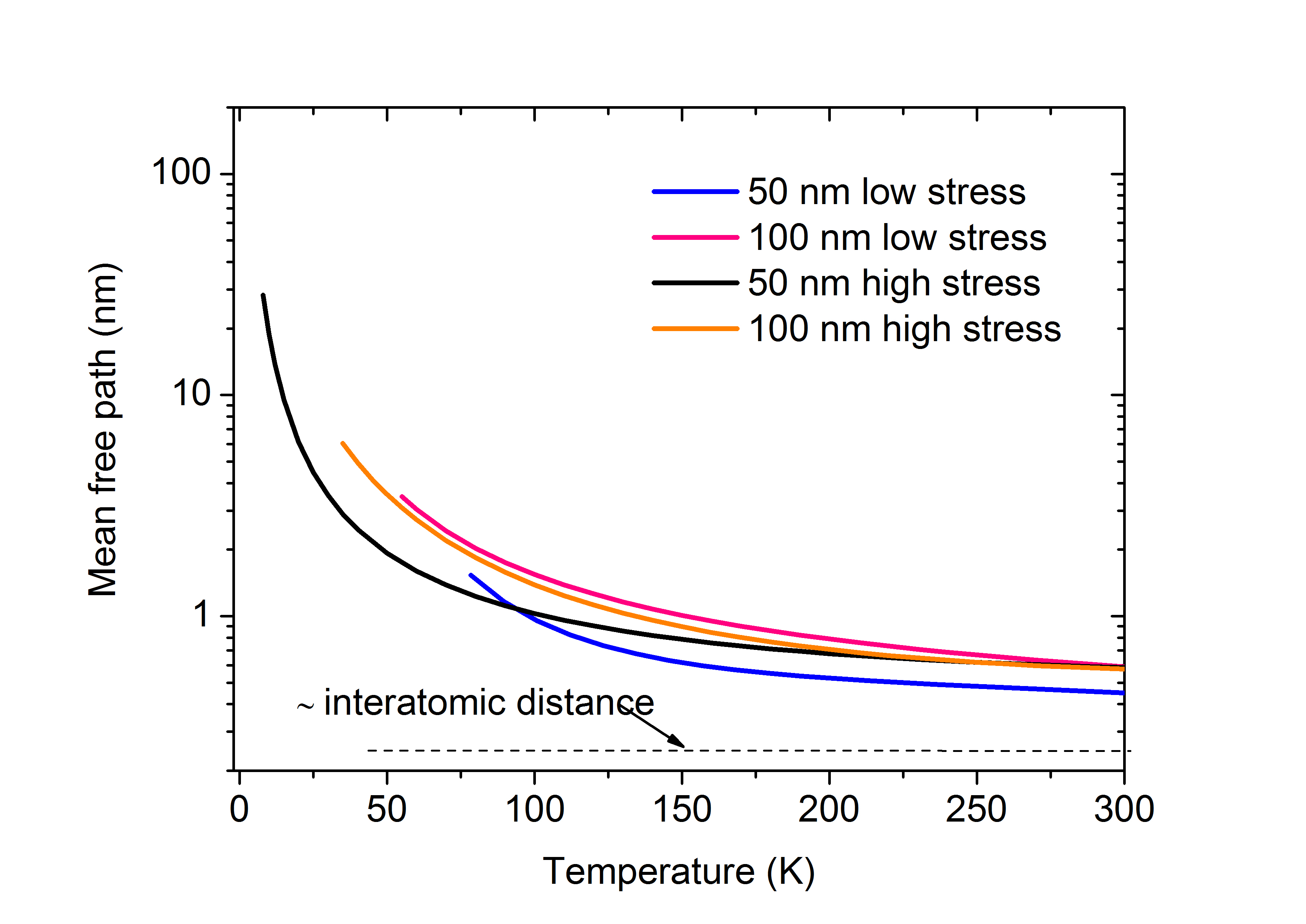} 
		\caption{(colour online) Mean free path $\Lambda$ of measured samples calculated using experimental data of specific heat and thermal conductivity. The dashed line shows the estimation of the mean free path using the Debye specific heat.}
		\label{mean}
	\end{center}
\end{figure}

In the theoretical work by Wu and Yu \cite{Wu2011}, the starting point is to consider that the stress (bond constraints, impurities, local defaults or even external strain) can modify either the TLS barrier height $V$ or the coupling between TLS and phonons denoted by $\gamma$. In this model, it is predicted that the modification of $V$ and $\gamma$ (by taking into account the amplitude of the stress in stoichiometric SiN) will have a significant effect on the thermal conductivity and mechanical dissipation. Let us discuss the two cases separately. First, when the barrier height is affected a difference between the thermal conductivity in low and high stress should be seen as the temperature is reduced with a factor close to five at 50~K. This is clearly not observed in our measurements since the thermal conductivity of the HS and LS membranes are very similar. We can only point out that around 50~K the thermal conductivity is slightly different between 50~nm and 100~nm samples, a behaviour that can be attributed to a reduction of mean free path in the thinner membrane. Secondly, in the case of the coupling between phonon and TLS (given by the parameter $\gamma$), an effect even larger is expected with a thermal conductivity a factor of ten higher in the HS SiN as compared to the LS at room temperature. This could be indeed very interesting for practical applications. Even though the stoichiometry is not strictly identical between the low and high stress membranes, we do not observe such a big difference in thermal conductivity. This has to be drawn closer to the mechanical measurement performed at 4~K, which also did not present any large differences in mechanical damping between HS and LS devices. 

We thus demonstrated negligible effect of stress on the thermal conductivity and mechanical dissipation in amorphous SiN.
We conclude that the hypothesis of TLS which barrier height $V$ or coupling constant $\gamma$ is affected by stress does not apply to these materials in the present stress range.
We also underline that the values of thermal conductivity we have measured for both high stress Si$_3$N$_4$ and low stress SiN membranes are in perfect accordance with most of the values already published (see table~\ref{tab:SiN}).
In order to highlight the low temperature particularities of the phonon conductivity in these thin membranes, it is particularly important to discuss the temperature variation of the mean free path \cite{Sultan2013}. Fig.~\ref{mean} shows the phonon mean free path in the membranes determined from the kinetic equation $\Lambda= 3k/Cv_s$, $v_s$ being the Debye speed of sound. It has been shown in the past that this equation can be used even at room temperature for amorphous materials by Pohl and co-workers \cite{Pohl2002}. At 300~K all curves (with the exception of 50~nm LS) approach the same limit which is two times higher than the inter-atomic spacing (0.25~nm for amorphous SiN). This is in very good agreement with previous thermal analysis \cite{Sultan2013}. As the temperature decreases, the mean free path increases rapidly to reach the order of ten nanometers at 20~K. As it can be seen in Fig.~\ref{mean}, it is reasonable to ascribe the difference of thermal transport below 200~K to a reduced mean free path in the thinner membranes.

\section{Conclusions}

The thermal conductivity has been measured on silicon nitride membranes having low and high stress. The objective was to search for any effect of internal stress on the phonon thermal conductivity and mechanical dissipation. Even though very high stress (of the order of 1~GPa) has been evidenced in suspended stoichiometric SiN membranes by nanomechanical measurements, it has been shown using very sensitive 3$\omega$ technique that the thermal conductivity was not affected.
Besides, mechanical dissipation is almost independent of stress, even though high $Q$s are obtained in HS structures in accordance with the "dissipation dilution" model.
This rules out a scenario of strong increase of thermal conductivity (and concomittantly a strond decrease of mechanical dissipation) with the presence of stress proposed recently by Wu and Yu \cite{Wu2011}, either through the increase of the barrier height of two level systems or through the decrease of the coupling between TLS and phonons.
We also show that the thermal properties of the most commonly used silicon nitride materials are equivalent. We then express doubts about the possible use of stress in thermal engineering in amorphous materials.

\section{Acknowledgments}
We acknowledge technical supports from Nanofab, the Cryogenic and the Electronic facilities and the Pole Capteur Thermom\'etrique et Calorim\'etrie of Institut N\'eel for these experiments. Funding for this project was provided by a grant from La R\'egion Rh\^one-Alpes (Cible and CMIRA), by the Agence Nationale de la Recherche (ANR) through the project QNM no. 0404 01, by the European projects: MicroKelvin FP7 low temperature infrastructure Grant no. 228464 and MERGING Grant no. 309150.

\end{document}